\PassOptionsToPackage{table}{xcolor} 
\documentclass[draft]{agujournal2019}
\usepackage{url}
\usepackage{lineno}
\usepackage[inline]{trackchanges}
\usepackage{soul}
\usepackage{amsmath}
\usepackage{longtable}   
\draftfalse
\journalname{Space Weather}

\begin{document}

\title{Machine Learning-Ready Data Sets for the Analysis and Nowcasting of Atmospheric Radiation at Aviation Altitudes}

\authors{V. M. Sadykov\affil{1}, Z. M. Watkins\affil{1}, D. Kempton\affil{1}, W. Jones\affil{1}, S. K C\affil{1}, G. T. Goodwin\affil{1}, X. He\affil{1}, W. K. Tobiska\affil{2}, I. Kitiashvili\affil{3}, C. Mertens\affil{4}, S. Ranjan\affil{3}, D. G. Deardorff\affil{5}, R. Spaulding\affil{5}}

\affiliation{1}{Georgia State University}
\affiliation{2}{Space Environment Technologies}
\affiliation{3}{NASA Ames Research Center}
\affiliation{4}{NASA Langley Research Center}
\affiliation{5}{InuTeq, LLC}

\correspondingauthor{Viacheslav M Sadykov}{vsadykov@gsu.edu}
\correspondingauthor{Zachary Mark Watkins}{zwatkins1@student.gsu.edu}

\begin{keypoints}
\item We present an open-access machine learning-ready dataset for nowcasting the atmospheric radiation environment based on a diverse set of Geospace properties.
\item Flight data points are split into three partitions with non-overlapping flights that equally sample the parameter space.
\item The nowcasting test case demonstrates results comparable to predictions of physics-based radiation environment models.
\end{keypoints}

\begin{abstract}
Nowcasting and forecasting of the radiation environment in the Earth’s lower atmosphere are critical for the safety of aircraft and spacecraft crews and passengers. Currently, this problem is addressed by employing statistical and physics-based models that take into account particle transport and precipitation. However, given the increased number of radiation measurements available to the community, it is possible to start developing data-driven approaches. We prepared Machine Learning-ready (ML-ready) datasets to nowcast the effective dose rates at aviation altitudes. The presented datasets contain 92,476 individual measurements from 589 flights obtained by the Automated Radiation Measurements for Aerospace Safety (ARMAS) experiment from 2013 to 2023. The ARMAS measurements are augmented with the properties of the Geospace environment, such as solar soft X-ray and proton fluxes, solar wind properties, secondary cosmic ray neutrons, space weather indexes, and global solar activity indicators (such as daily sunspot number). ARMAS data are separated into three partitions, ensuring that (1) the data points from a single flight remain within the same partition, and (2) each partition samples the flight locations and Geospace environment conditions equally. Several versions of the datasets allow predictions based on point-in-time measurements and use up to 24 hours of Geospace parameter history. The test of the use case demonstrates a possibility of nowcasting ARMAS measurements with accuracies slightly better than the considered physics-based models. The publicly available ML-ready datasets could serve as the first step in data preparation for ML-driven nowcasting and forecasting of the radiation environment.
\end{abstract}

\section*{Plain Language Summary}

Understanding the radiation levels in the Earth's atmosphere and predicting them at the locations along flight routes is important for the aviation industry. Machine Learning (ML) techniques are often used nowadays for prediction tasks. However, an extensive data preparation phase is required before applying an ML algorithm for predictions. Given the advantage of increasing volumes of the radiation measurement data, we present ML-ready datasets that allow users to bypass the data preparation stage and go directly to the phase of testing ML models for aviation radiation prediction. The ML-ready dataset is publicly available via the Radiation Data Portal (\url{https://dmlab.cs.gsu.edu/rdp/ml-dataset.html}).

\section{Introduction}

Understanding the radiation environment at the aviation altitudes (approximately 8 -- 17\,km) remains an important and not fully explored topic for the aviation community. Integrating the effective dose rate along the flight trajectory constitutes the total effective dose absorbed by the aircraft crew and passengers. Given that there are recommended effective dose limits of 20\,mSv per year averaged over 5 years (a total of 100\,mSv in 5 years) for radiation workers, and 1\,mSv per year for general public \cite{cho2017radrates}, accurate monitoring and prediction of the radiation environment becomes essential.

Galactic cosmic rays cascading in Earth's atmosphere are the primary contributors to effective dose rates. The atmospheric radiation dose rates can also increase during strong solar energetic particle (SEP) events \cite{Reames2021}. For example, the estimates presented for the September 10-11, 2017, SEP event demonstrated that the location-averaged effective dose rate due to the SEP at a height of 12\,km was approximately 3\,$\mu{}Sv/h$ \cite{Kataoka2018}. While this is less than the contribution due to galactic cosmic rays (more than twice that amount), it becomes comparable by an order of magnitude. In addition, the recent works indicate that the radiation dose rates can be enhanced due to the plasmaspheric hiss wave activity \cite{Aryan2023SpWea..2103477A,Aryan2025JGRA..13033959A} and related enhanced precipitation of electrons into the upper atmosphere, providing even deeper connection of the radiation measurements at aviation altitudes to the near-Earth space environment.

Routine measurements of the radiation environment have become possible in the last decade via efforts from the Automated Radiation Measurements for Aerospace Safety experiment \cite<ARMAS;>[]{Tobiska2015ARMAS,Tobiska2016ARMAS,Tobiska2018ARMAS}. The ARMAS device measures the local radiation environment conditions in real time during commercial aircraft flights. Currently, ARMAS has flown on more than 1000 flights and provided $\sim$400,000 individual measurements, sampling the aviation altitudes across the United States and worldwide (\url{https://dmlab.cs.gsu.edu/rdp/}). The development and release of the Radiation Data Portal \cite<RDP;>[]{Sadykov2021RDP} provided a convenient overview and access to the  ARMAS data.

Nowcasting or predicting the radiation environment in Earth's atmosphere can be carried out by physics-based approaches such as CARI \cite<i.e., the Civil Aviation Research Institute;>[]{copeland2021cari7}, Professional AviatioN DOse CAlculator \cite<PANDOCA;>[]{Matthia2014SpWea..12..161M}, the Nowcast of Aerospace Ionizing RAdiation System \cite<NAIRAS;>[]{Mertens2013}, etc. The NAIRAS model takes into account the galactic cosmic ray and SEP inputs. It is closely integrated with the ARMAS experiment by providing the assessments of the radiation dose rates for every ARMAS measurement. Recently, version 3 of the NAIRAS model was released \cite{Mertens2023,Mertens2024} that integrates the barometric altitude corrections for a better comparison to ARMAS measurements at the appropriate altitudes, as well as run-on-request capabilities at the Community Coordinated Modeling Center (CCMC) by the National Aeronautics and Space Administration (NASA). In addition, with the now substantial amount of available ARMAS data, it has become possible to explore data-driven, statistical, and machine learning (ML) based analysis for prediction conditions of the radiation environment. A comparison of data-driven and physics-based predictions is valuable for further improvements to the physics considered in the models.

Data preparation is the first, yet most demanding step when developing an ML model. Typically, datasets that can be easily fed into ML models with relatively low or no preparation required are called ML-ready datasets. Several works \cite{Masson2024,Nita2022} have introduced the key principles required for these datasets, often related to the overall completeness, preparation levels, and accessibility to a wider community. The preparation of the ML-ready datasets suitable for radiation environment prediction could enhance community efforts in modeling the radiation environment and stimulate the development of data-driven approaches.

The primary goal of this paper is to present and provide a detailed description for the ML-ready dataset that could be used for nowcasting and forecasting of the radiation environment at aviation altitudes, and support it with the use-case example. The paper is structured as follows: Section~\ref{sec:data} describes the data mining and preprocessing steps required for both ARMAS and the space environment measurements. Section~\ref{sec:partition} presents steps for shaping the data into an ML-ready format. This includes the association of ARMAS measurements with the Geospace environment properties (such as ground-based neutron monitor counts, soft X-ray and proton flux measurements at the geostationary orbit, solar wind properties at L1 point, geomagnetic and global solar activity indexes) and the partitioning of multi-dimensional sparse data into three subsets that should be used directly for training and validation of the model. Section~\ref{sec:example} provides a case example of how the dataset could be used for nowcasting, followed by a summary and discussion in Section~\ref{sec:summary}.

\section{Data Preparation}\label{sec:data}

\subsection{ARMAS data processing}

\begin{figure}
    \includegraphics[width=1.0\linewidth]{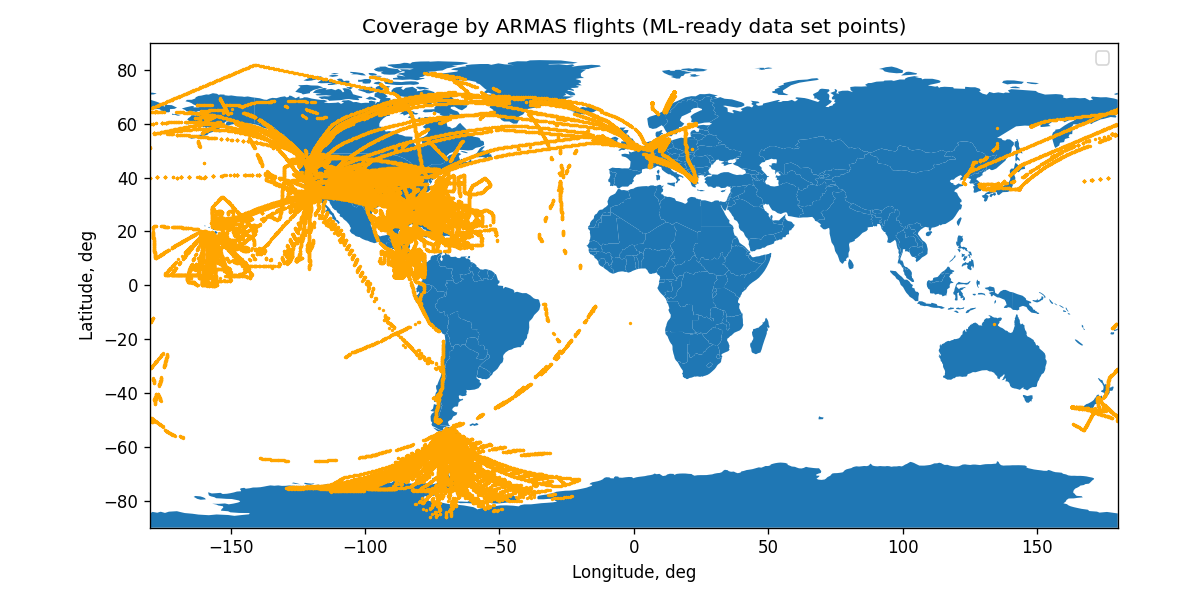}
    \caption{Coverage of considered ARMAS flight measurements over the Earth globe. The individual measurements are marked as orange points on the map.}
    \label{fig:armascoverage}
\end{figure}

The Automated Radiation Measurements for Aerospace Safety (ARMAS) experiment provides the richest publicly available dataset of atmospheric radiation measurements along routes at aviation altitudes. In this work, we consider ARMAS data obtained from 1142 flight files (589 of which passed the selection criteria and contributed to an ML-ready dataset) from June 2013 to December 2023. This period of time covers the peak and decay of solar cycle 24, the solar minimum, and the rise phase of solar cycle 25, which enables sampling different levels of global solar activity. The flights cover a wide geographic span, including the extensive sampling of the continental US and territories, the Pacific Ocean regions, the North Atlantic, and Antarctica (Figure~\ref{fig:armascoverage}). ARMAS measurements are stored in individual files for each flight and device and are accessible either via the Space Environment Technologies (SET) public data archive (\url{https://sol.spacenvironment.net/ARMAS/Archive/}) or the Radiation Data Portal (\url{https://data.nas.nasa.gov/helio/portals/rdp/}). More details on ARMAS measurements can be found in \citeA{Tobiska2015ARMAS,Tobiska2016ARMAS,Tobiska2018ARMAS}.

For each flight, we apply the following pre-processing steps. We sample the ARMAS measurements only in the barometric altitude range of 8.0\,km\,--\,15.5\,km, as ARMAS often measures zero radiation below these altitudes. The corresponding maximum GPS altitude in the filtered dataset is 16.272\,km. We notice that several flights had two or more ARMAS devices onboard, and the measurements of these devices were reported in separate files. For groups of files corresponding to the same flight, we only consider the single file where ARMAS measurements are most strongly correlated with the NAIRAS v3 model estimates (in terms of Pearson correlation coefficient). Additionally, we remove the data points marked for non-science-use, those during periods of electromagnetic interference, and those with unphysically high dose rate measurements ($>$50\,$\mu$Sv/h). The remaining data points are inspected manually for every flight. The procedure results in 589 unique flights propagating into the final dataset, and 92,476 unique data points.

The ARMAS files prepared by the SET public data archive (\url{https://sol.spacenvironment.net/ARMAS_Archive/}) contain several additional parameters that we propagate into the final dataset. For example, there is information about the geomagnetic cutoff rigidities \cite<provided as a courtesy of SSSRC,>[]{SmartShea2009AdSpR..44.1107S} as well as the barometric and GPS altitudes. The files also contain the physics-based nowcast of the radiation dose rate obtained with 2nd and 3rd versions of the Nowcast of Atmospheric Ionizing Radiation for Aviation Safety model \cite<NAIRAS;>[]{Mertens2013,Mertens2023,Mertens2024}. Version 3 (v3) of the model provides a better agreement with the ARMAS measurements and is therefore suggested to be used as a physics baseline for the radiation forecasts.

\subsection{Neutron monitor stations}\label{sec:data:nm}

Along with the muons, secondary cosmic ray neutrons are an abundant remnant of the atmospheric particle cascades formed by the primary cosmic rays and, therefore, represent one of the most reliable ways to monitor the Geospace cosmic ray environment. The Neutron Monitor Database \cite<NMDB,>[]{Kozlov2003ESASP.535..675K,Vaisanen2021JGRA..12628941V} agglomerates the measurements from more than 50 neutron monitors around the globe. The NMDB Event Search Tool (NEST, \url{https://www.nmdb.eu/nest/}) provides access to the data from these stations, including the count rates corrected with respect to the effects of the local atmospheric pressure.

The NAIRAS physics-based model utilizes neutron monitor data from four locations, Oulu (\texttt{OULU}, $R_c$=0.81\,GeV), Lomnicky (\texttt{LKMS}, $R_c$=3.84\,GeV), Thule (\texttt{THUL}, $R_c$=0.30\,GeV), and Izmiran (\texttt{MOSC}, $R_c$ = 2.43\,GeV), as inputs. However, according to the NEST interface, \texttt{LKMS} and \texttt{MOSC} stations have some interruptions for the period of interest (2013-2023). The continuity of data is critical for the construction of the ML-ready dataset. Therefore, we decided to replace these stations with the neutron monitor station in Newark (\texttt{NEWK}, $R_c$=2.40\,GeV), and we also added the South Pole station (\texttt{SOPO}) with extremely low geomagnetic cut-off rigidity ($R_c$=0.10\,GeV). The pressure- and efficiency-corrected counts are obtained for all stations with a cadence of 5 minutes. For the time intervals in which the measurements of the \texttt{THUL} station exceeded 300 counts/s (unrealistically-high values for this station), the linear interpolation has been applied. The same procedure was used for the \texttt{NEWK} station, using a threshold of 120 counts/s. These thresholds were adopted through visual inspection of the flux dynamics over the 11-year period of interest.

\subsection{Solar wind parameters}

According to \citeA{Tobiska2018ARMAS}, ARMAS measurements tend to have higher radiation dose rates than the predictions by the NAIRAS physics-based models, specifically in the regions of low geomagnetic cutoff rigidity. The recent studies \cite{Aryan2023SpWea..2103477A,Aryan2025JGRA..13033959A} find a strong correlation between the enhanced dose rates observed by ARMAS and plasmaspheric hiss waves enhancing precipitation of the electrons from the Van Allen radiation belts into the upper atmosphere. Given that the perturbations of the Earth's magnetosphere and the dynamics of the radiation belts are tightly coupled with the solar wind parameters, we include the properties of the solar wind in our dataset. The curated solar wind parameters are accessible via the OMNIWeb (\url{https://omniweb.gsfc.nasa.gov/}) online dataset with a 5-minute cadence. We include all major properties, such as the solar wind density, temperature, the magnitude, and all three components of the velocities and magnetic fields. The linear interpolation has been applied for the time periods when the solar wind properties were not available or the placeholder values (like `9999' for the magnetic field) have been encountered. All data from OMNIWeb is checked for continuity, and no affecting data gaps were found.

\subsection{Energetic particles}

Radiation levels at aviation altitudes might significantly increase (even exceed the galactic cosmic ray component) during solar energetic particle events \cite{Kataoka2018}. \citeA{Tobiska2018ARMAS} also noted that there are many enhanced radiation events in ARMAS data obtained for L-shells between 1.5 and 5 in the Western hemisphere, and indicated the radiation belt particles as a possible reason for this. Therefore, we incorporate the energetic proton and electron measurements in our database, all measured by the Geostationary Operational Environmental Satellite (GOES) series. The proton fluxes are obtained in 7 integrated energy channels: $\geq$1\,MeV, $\geq$5\,MeV, $\geq$10\,MeV, $\geq$30\,MeV, $\geq$50\,MeV, $\geq$60\,MeV, and $\geq$100\,MeV. The data are obtained from the Integrated Space Weather Analysis (ISWA) webapp (\url{https://ccmc.gsfc.nasa.gov/tools/ISWA/}). The energetic electron flux measurements are included only in one energy channel, $\geq$2\,MeV, because of the data continuity issues found in other energy channels. The resolution of both the proton and electron particle data utilized for the dataset construction is is 5\,min. For the time periods when the particle flux in a certain channel has dropped below to zero, the linear interpolation has been applied in this channel.

\subsection{Solar X-ray measurements}

Solar soft X-ray emission integrated over the solar disk is traditionally measured by the GOES spacecraft series, using the onboard X-ray sensor. We are utilizing 1-minute averaged fluxes in two channels, 1\,--\,8\,\AA{} and 0.5\,--\,4\,\AA, as input to our dataset. A similar type/cadence of the data was incorporated in the currently available Radiation Data Portal \cite{Sadykov2021RDP}. For the time periods when the flux in 1\,--\,8\,\AA{} or 0.5\,--\,4\,\AA channel has dropped below $2\times{}10^{-9}$\,W$\cdot{}m^{-2}$, the linear interpolation has been applied in the corresponding channel.

\subsection{Geomagnetic activity indexes}

We adopt hourly planetary Kp, Ap, and Dst indexes, all available via the OMNIWeb (\url{https://omniweb.gsfc.nasa.gov/}) online dataset with a 1-hour cadence.
The presented indices reflect the state of the Earth's magnetosphere during the major interplanetary disturbances, such as the presence of the Interplanetary Coronal Mass Ejections (ICMEs) or high-speed stream interaction regions in the solar wind. The variations in these parameters can affect the geomagnetic cutoff rigidity \cite{Ptitsyna2021Ge&Ae..61..169P} and, therefore, modulate the galactic cosmic ray precipitation into the Earth's atmosphere.

\subsection{Global solar activity and characteristics}

In addition to the transient activity effects of the Sun on the Geospace, there are longer-term effects associated with the global solar activity. Although the neutron monitor data can capture effects related to the global solar activity (Section~\ref{sec:data:nm}), the solar activity indexes make a picture of the evolution of the radiation environment more complete. For example, \citeA{Koldobskiy2022SoPh..297...38K} found the delay of about 7 months of the sunspot numbers with respect to the neutron monitor measurements. Therefore, we include the following global solar activity indicators:

\begin{itemize}
    \item {\bf Solar F10.7 index.} The Solar F10.7~index, which measures solar radio flux at a wavelength of 10.7\,cm (2800\,MHz), is a critical indicator of solar activity and has significant implications for Earth's atmospheric conditions. F10.7 index is accessible via the OMNIWeb (\url{https://omniweb.gsfc.nasa.gov/}) online dataset with a 1-hour cadence. For the time intervals when the F10.7 flux had a placeholder value of `999', the linear interpolation has been performed.
    \item {\bf Daily sunspot number}, which tracks the daily count of visible sunspots on the Sun's surface, is a traditional characteristic of solar activity level. Managed by the World Data Center SILSO (Sunspot Index and Long-term Solar Observations, \url{https://www.sidc.be/SILSO/datafiles}), hosted at the Royal Observatory of Belgium, these data provide critical data for understanding solar cycles and their impacts on space weather. For the days when the daily sunspot number was not provided, the linear interpolation has been applied.
    \item {\bf Solar polar fields.} The solar polar magnetic fields, monitored by the Wilcox Solar Observatory (WSO, \url{http://wso.stanford.edu/Polar.html}) at Stanford University, are crucial indicators of the Sun's magnetic activity and the solar cycle's progression. These fields, located at the Sun's poles, reverse approximately every 11 years, marking a new solar cycle. The strength and structure of the polar fields are key to predicting the magnitude of future solar cycles, as they play a central role in the generation of the Sun’s magnetic field through the solar dynamo process. The WSO has provided near-continuous, detailed measurements of solar polar magnetic fields since 1976. It is an invaluable resource for understanding solar dynamics, space weather prediction, and their influence on the heliosphere. For the periods when the solar polar fields were provided, the linear interpolation has been applied.
\end{itemize}

\begin{figure}
    \includegraphics[width=1.0\linewidth]{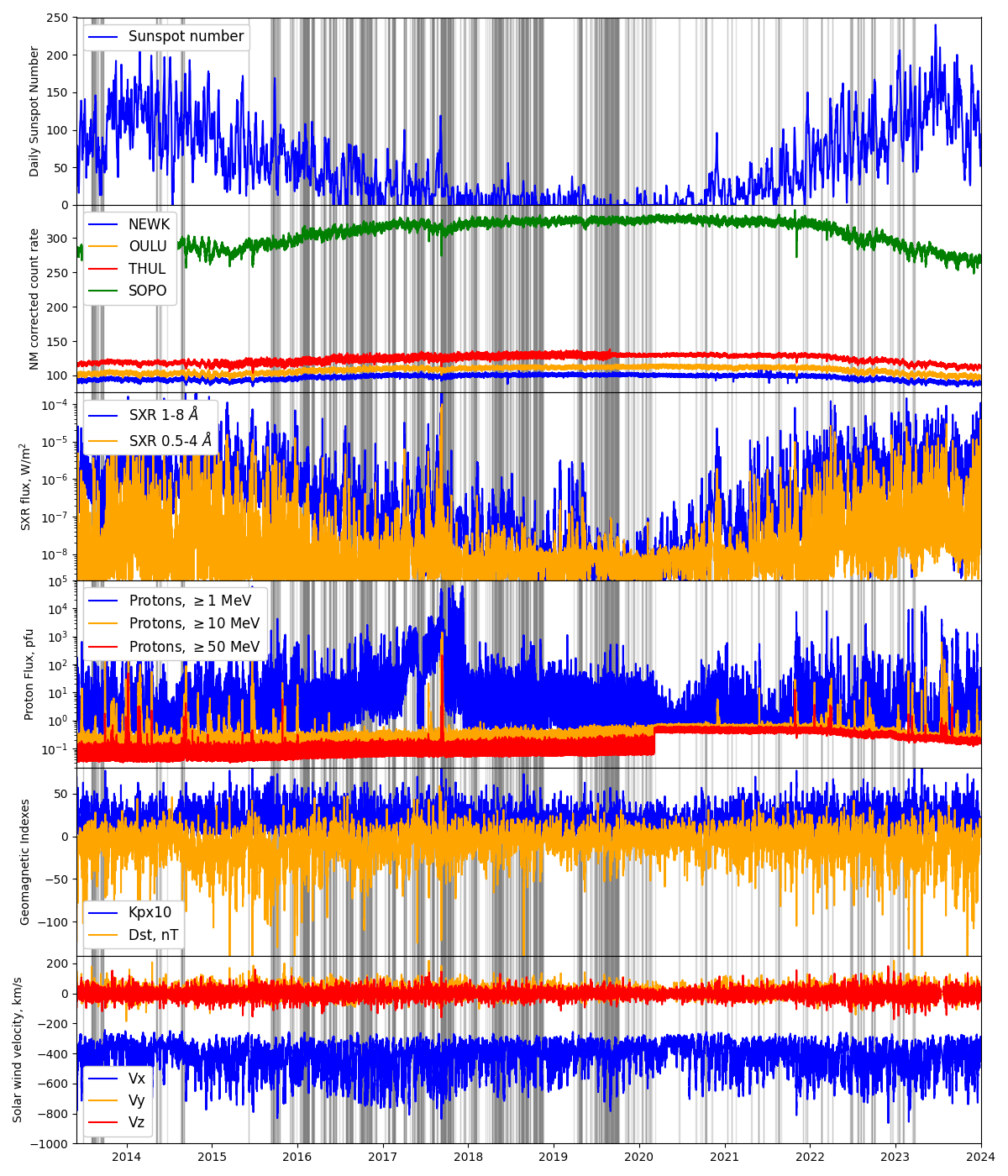}
    \caption{Evolution of some selected Geospace environment parameters from June 2013 until December 2024, from top to bottom: daily sunspot number, neutron monitor corrected counts from four stations considered, integrated soft X-ray fluxes, energetic proton fluxes, geomagnetic indexes (Kp and Dst), and solar wind velocities at L1. Gray lines in the background represent the time moments covered by the ARMAS flight measurements.}
    \label{fig:environment_vis}
\end{figure}

The illustration of some of these sources is presented in Figure~\ref{fig:environment_vis}. As one can see, most features are continuous across the time interval of interest. The only noticeable change occurs in high-energy proton flux ($\geq$50\,MeV) with a jump in the background. However, since the protons of only a very high energy and flux could impact the radiation levels, the jumps in the proton flux background levels are not a point of concern.

\section{ML-Ready Data Set Construction}\label{sec:partition}

Although ARMAS data and the corresponding Geospace environment characteristics have been collected and cleaned for the time interval of interest (June 2013~-- December 2023), the data is still not ready for ML purposes. As specified in \cite{Nita2022,Masson2024}, while there is no specific criteria to classify a particular dataset as ML-ready, these datasets typically feature the following properties: Accessibility (the data should be accessible for the users) , Completeness and Integrability (all pieces of data should be available to the users at no extra effort), volume Sufficiency (large enough data points for machine learning models), Cleanliness (data pre-processed and inspected), and Understandability (accessible for non-domain experts). Our ML-ready dataset must feature additional characteristics. Specifically, the dataset should contain all components necessary for training/evaluating the ML models and summarizing/integrating the results, and should not require additional domain knowledge needed before ML modeling (such as related to data partitioning). Therefore, at least two more steps are required: the ARMAS measurements should be merged with the preceding environmental parameters, and the dataset has to be partitioned. To support the understandability of the data, partitions should be created with input from domain experts to prevent artificial correlations.

\begin{figure}
    \includegraphics[width=1.0\linewidth]{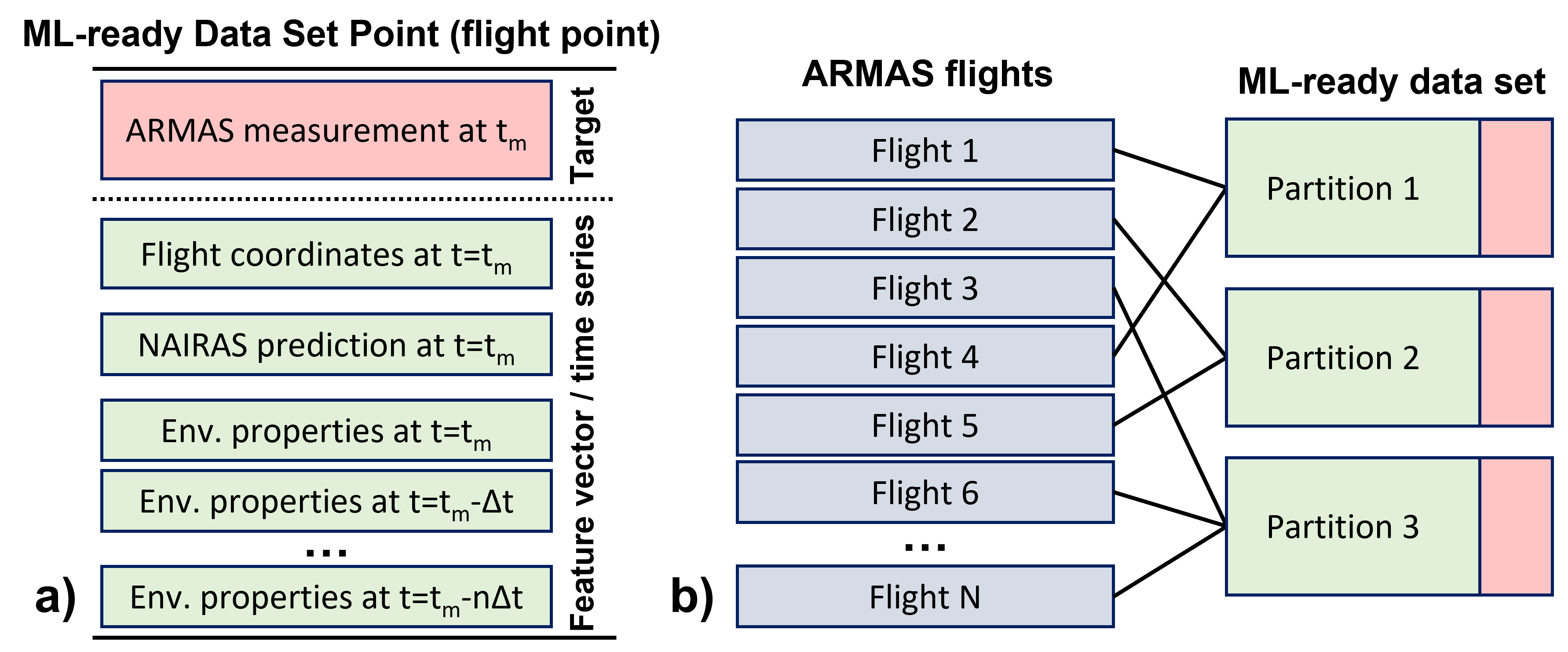}
    \caption{(a) Schematic structure of the ML-ready dataset entity. A target corresponds to the measurement of radiation dose rate during the ARMAS flight, and the feature vector represents the flight timing and coordinates, NAIRAS predictions, and prehistory of the measurements of the environment. (b) Illustration of the subdivision of ARMAS flights into partitions.}
    \label{fig:partitions_schema}
\end{figure}

We associate individual ARMAS measurements with the temporally closest preceding measurements of the Geospace environment. We avoid using any temporal interpolations to avoid breaking the causality principle and ensure that we are not utilizing information from times in the future for the radiation nowcasting (Figure~\ref{fig:partitions_schema}, left). Given that some parameters, like the daily sunspot numbers, have a relatively large time cadence (24 hours in this case), the preceding measurement of the daily sunspot number may represent the conditions up to 24 hours before the actual measurements by ARMAS.

The second step is the separation of the individual ARMAS measurements into the partitions (Figure~\ref{fig:partitions_schema}, right). This process indicates the chunks of data that can be used for training, validation, and testing. This ensures the reproducibility of the research and the direct comparison between different models if trained on the same partitions. The partitioning previously was used in other ML-ready datasets, such as the Space Weather Analytics for Solar Flares, SWAN-SF \cite{Angryk2020NatSD...7..227A}. The key point is that the ARMAS measurements cannot be separated randomly. The data points from the same flight could be just one minute away from each other. Therefore, the environmental properties are similar. Thus, if these data are split between the train and test partitions, this could generate artificial correlation between the partitions. This effect was previously recognized as a `temporal coherence' in flare forecasting \cite{Ahmadzadeh2021ApJS..254...23A}. Therefore, the data from every ARMAS flight must be allocated to the same partition.

In principle, each partition should represent the entire dataset, meaning the distribution of the parameters in each partition should represent the entire dataset \cite{Liu2019partitioning}. Separation into partitions could be a challenging problem if data coverage is sparse, and the problem is multi-dimensional. Unfortunately, this is the case for the considered dataset: the measurements are acquired along slightly more than 1000 individual flight trajectories, and more than 40 Geospace parameters are associated with every measurement. To overcome the manual trial-and-error partitioning process, we utilized an ML clustering algorithm to help with it. The steps are as follows:

\begin{enumerate}
    \item \textbf{Down-selection of the five parameters for clustering.} As was indicated above, the number of Geospace parameters is relatively large. Distance-based clustering algorithms struggle with high dimensionality (the so-called `curse of dimensionality' problem). However, many of the parameters are expected to be correlated with each other. For example, there is a strong correlation between all geomagnetic indices (such as Kp, Ap, Dst) and solar wind parameters at L1 point, global solar activity parameters (daily sunspot numbers, F10.7 flux, polar magnetic field measurements, neutron monitor measurements), description of the location in geomagnetic coordinates (geomagnetic latitude, L-shell, cutoff rigidity), etc. Instead of considering all of them, we down-select five parameters based on which we will cluster the data. The first three are related to the location of the measurement: geomagnetic longitude, geomagnetic latitude, and barometric flight altitude. The fourth parameter describes the geomagnetic activity, for which we use the Dst index. The Dst index was chosen among all geomagnetic indices because its distribution is closest to a normal distribution and shows the least skewness. The last parameter we use is the daily sunspot number that reflects global solar activity levels. Although this selection is not unique, it is sufficient for partitioning purposes. Because the number of ARMAS flights during the solar energetic particle (SEP) events is very low, we did not use the SEP-related measurements for partitioning.
    \item \textbf{Clustering of individual measurements.} We apply the Gaussian Mixture Model (GMM) clustering based on the five parameters for the individual ARMAS measurements. The number of GMM components has to be sufficiently large to create a relatively equal representation for all parameters in every partition. We note here that the GMM is a soft clustering methodology as it assigns the probabilities of every point to belong to a certain cluster rather than associating it with a particular cluster.
    \item \textbf{Associating flights with clusters.} Each individual measurement in each flight now has a probability (or a weight) of belonging to each of the clusters. To associate the entire flight with the particular cluster, we sum up the probabilities of the individual measurements in this flight for every cluster. Therefore, a flight becomes assigned entirely to a particular cluster for which the sum of the probabilities of the individual measurements is the highest. We perform these assignments for all flights in our dataset.
    \item \textbf{Separation of the flights into three partitions.} We start from the first cluster and distribute the flights to this cluster into three partitions in sequential order. After distributing all flights in the cluster, we move to the next cluster and repeat the procedure starting from the next partition. For example, if cluster 1 has four flights, then flight \#1 will be distributed to partition 1, flight \#2~-- to partition 2, flight \#3~-- to partition 3, and flight \#4~-- again to partition 1. Then one moves to cluster 2, and flight \#1 in this cluster goes now to partition 2, flight \#2~-- to partition 3, etc. This procedure ensures that all partitions will include flights belonging to the same cluster (and, therefore, likely sampling similar spatial locations and geomagnetic and solar activity levels).
\end{enumerate}

The number of the GMM components/clusters is one of the parameters of the partitioning algorithm. To understand the optimal number of clusters, we have implemented the partitioning for different cluster numbers $n_{cl}$ = $\{1, 2, 5, 10, 15, 50, 100, 200\}$ and quantified the similarity of the parameter distributions across the partitions. The first approach for the similarity quantification is the Kullback–Leibler Divergence (KL Divergence) which is defined for any two probability distributions $p(x)$ and $q(x)$ as:
\begin{gather}
    D_{KL}(p||q) = \int p(x) log \dfrac{p(x)}{q(x)}dx
\end{gather}
Since $D_{KL}(p||q)\neq{}D_{KL}(q||p)$, we compute the average KL Divergence over the six possible ordered partition pairs for each parameter and the cluster number case. We then normalized the average KL Divergences to the average KL Divergence of the random partitioning (when only 1 cluster is chosen) and averaged again over the five parameters on which the clustering has been performed. The second approach that we considered is quantification of the spread of standard deviations (widths) of the distribution. For each number of clusters and each parameter in each partition, we have computed the width of the distribution as its standard deviation. We then quantified the spread of the widths across the partitions for each parameter as a standard deviation of widths, normalized it to the spread of the random partitioning, and then averaged over the five parameters on which the clustering has been performed.

\begin{figure}
    \centering
    \includegraphics[width=0.8\linewidth]{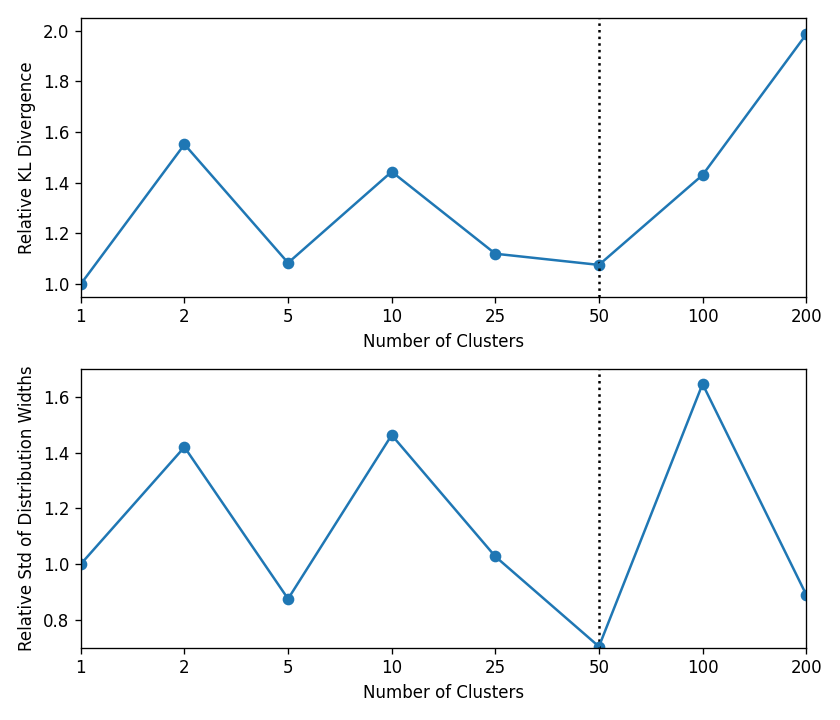}
    \caption{Distribution of the relative KL divergence and the relative standard deviation of the distribution widths for the distributions across three dataset partitions as a function of the number of GMM clusters. The vertical dashed line displays the number of clusters selected for the final dataset construction.}
    \label{fig:cluster_analysis}
\end{figure}

The results of this analysis are presented in Figure~\ref{fig:cluster_analysis}. As one can see from the top panel, interestingly, the random partitioning results in the most optimal KL Divergence. The next most optimal number of clusters is $n_{cl}=50$, which corresponds to the relative KL Divergence of $\sim$1.075. However, the bottom panel of the Figure~\ref{fig:cluster_analysis} illustrates that the partitions agree much better on the parameter distribution widths in case of $n_{cl}=50$ with respect to the random partitioning: the corresponding parameter-averaged relative standard deviation of the widths is $\approx$0.704. Therefore, for the final version of the dataset, we have chosen $n_{cl}=50$ for the GMM clustering.

\begin{figure}
    \includegraphics[width=1.0\linewidth]{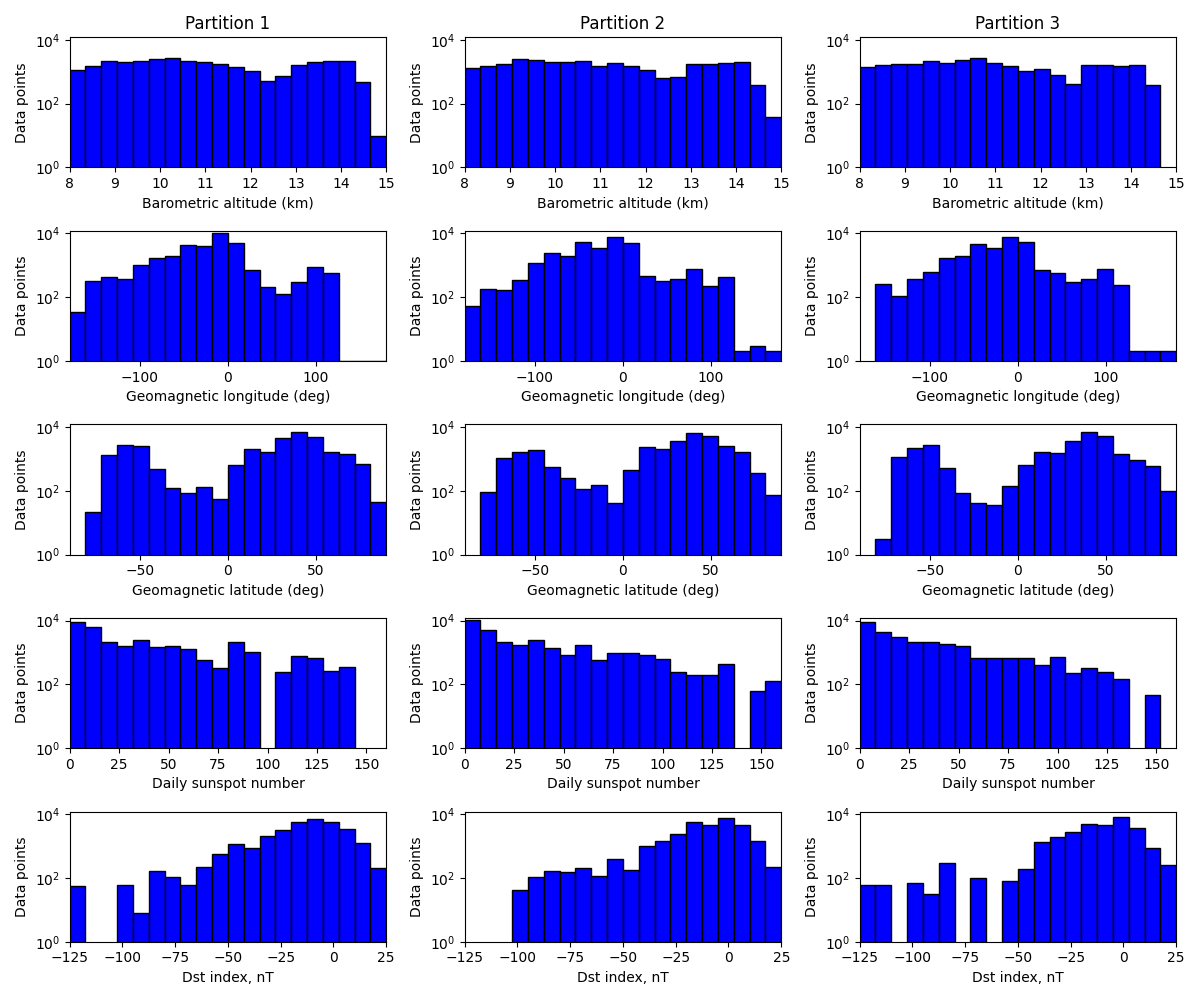}
    \caption{Distribution of the parameters used for clustering the data points (barometric altitude, geomagnetic longitude, geomagnetic latitude, daily sunspot number, and Dst index) within each partition of the dataset. Each row corresponds to a single parameter. The partition is indicated in the header of each column.}
    \label{fig:partitions_primary}
\end{figure}

While the partitioning strategy does not necessarily result in the `most optimal' distribution of the flights, it leads to a satisfactory representation of parameter combinations in every partition while avoiding the brute-force partitioning. The result of the distribution of the parameters within each partition is presented in Figure~\ref{fig:partitions_primary}. The histograms of the parameters are relatively similar for every column of the partitioning, indicating that each partition samples the parameter space relatively equally. Some differences are evident only for the extreme values of the parameters, such as the Dst indexes at around $\sim$ -100\,nT or the daily sunspot numbers of $\sim$150, which occur because not too many ARMAS flights have occurred during such high activity levels.
    
\begin{figure}
    \includegraphics[width=1.0\linewidth]{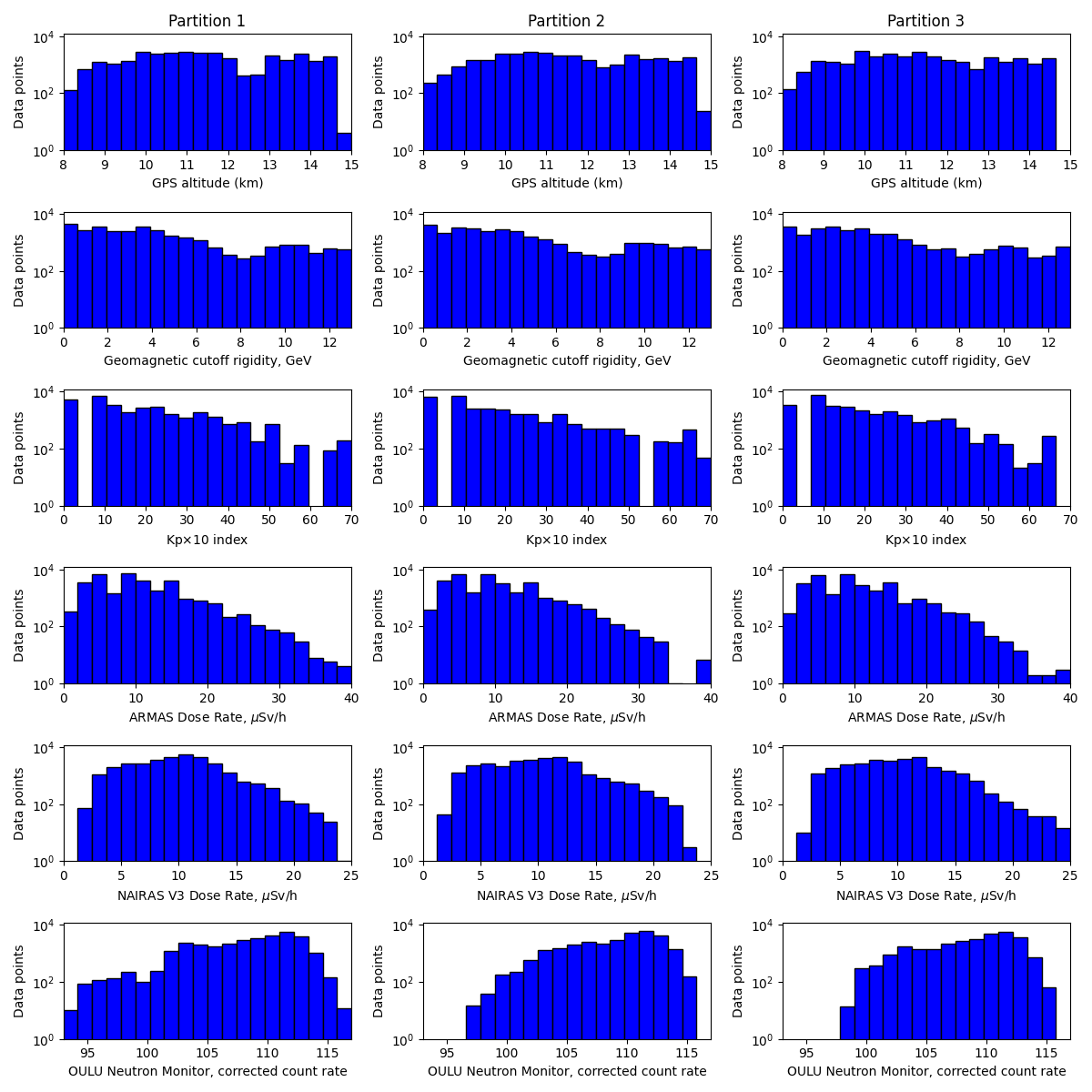}
    \caption{Distribution of the parameters that were not used for the clustering of the data points (GPS altitude, geomagnetic cutoff rigidity, Kp index multiplied by 10, ARMAS dose rate, NAIRAS dose rate, and corrected count rates of secondary cosmic ray neutrons detected by \texttt{OULU} station) within each partition of the dataset. Each row corresponds to a single parameter. The partition is indicated in the header of each column.}
    \label{fig:partitions_secondary}
\end{figure}

It is important to check if other parameters are sampled evenly across the partitions. Figure~\ref{fig:partitions_secondary} illustrates the distributions of the parameters in three partitions that have not directly participated in the clustering. The distribution of parameters is very similar, indicating a successful partitioning process.
    
The resulting data represents the ML-ready dataset that utilizes the last `snapshot' of Geospace properties before the ARMAS measurement. The ML-ready dataset consists of a feature vector (a set of characteristics based on which the prediction is made, representing Geospace parameters) and a target (a characteristic to predict, here ARMAS measurement). The feature vector comprises the flight timing and location properties, as well as the most recent properties of the environment. In addition to the most recent properties (Figure~\ref{fig:partitions_schema}), one can provide the time series of the evolution of the Geospace parameters up to the time before the measurements.  The preceding properties of the environment can be forward-interpolated with the cadence $\Delta{}t$ for $n$ time steps before the ARMAS measurement. At the same time, the partitioning of the individual ARMAS measurements and related time series remains as described above. Following this, we have finally constructed three publicly available ML-ready datasets:

\begin{itemize}
    \item The dataset that represents the most recent Geospace measurements and does not involve their time series (`static' dataset, $n=0$);
    \item The dataset that includes a 1-hour prehistory of the Geospace measurements before ARMAS flight measurement (`dynamic' dataset 1, $n=12$ and $\Delta{}t=5$\,min);
    \item The dataset that includes a 24-hour prehistory of the Geospace measurements before ARMAS flight measurement (`dynamic' dataset 2, $n=24$ and $\Delta{}t=1$\,hour);
\end{itemize}

All three versions of the datasets are currently accessible via the Radiation Data Portal (\url{https://dmlab.cs.gsu.edu/rdp/ml-dataset.html}).

\section{Dataset Use Case Example}\label{sec:example}

In this section, we illustrate how the constructed datasets can be utilized for ML-driven forecasting of atmospheric radiation. For this demonstration, we proceed with the simplest version of the three datasets constructed and described above, the `static' ML-ready dataset, which includes only the latest point-in-time measurement of every Geospace parameter. Among the ML approaches available off-the-shelf, we select the Random Forest regressor \cite{Breiman2001MachL..45....5B}, a bagging tree-based ensemble learning algorithm. Random Forest has previously been successfully applied to a variety of classification and regression problems in the space physics domain, including the prediction of solar flares and solar energetic particle events \cite{Liu2017ApJ...843..104L,OKeefe2022zndo...6780972O}, forecasting the duration of enhanced soft X-ray radiation during the flare \cite{Reep2021SpWea..1902754R}, the timing of the solar wind propagation \cite{Baumann2021JSWSC..11...41B}, ion-kinetic instability detection in the solar wind-like plasmas \cite{Sadykov2025arXiv250518271S}, etc. We note here that, despite the promising track record, one cannot guarantee that the Random Forest approach is the most optimal for the considered problem. Therefore, the survey of other ML models is highly encouraged and is one of the authors' goals for the future.

We have used the Random Forest model available at the {\tt scikit-learn} Python library package \cite{scikit-learn}. The model has several hyperparameters to optimize, such as the number of individual decision trees in the ensemble, the maximum depth restriction for each tree, the number of features randomly selected and propagated into every tree, etc. Typically, the hyperparameters are fine-tuned during the cross-validation phase when the models of different hyperparameters are evaluated on a designated partition or a subset of the training partition. Here, we do not perform the detailed cross-validation but rather use the parameters we found to perform satisfactorily for the considered study during our preliminary tests. Our Random Forest consists of 100 decision trees of a depth of 10 or less, with at least 2 samples from the dataset required for the split within the tree, and at least 4 samples to be at the leaf node. To guide the training process, we utilize the mean squared error as a measure of the regressor's performance.

\begin{figure}
    \includegraphics[width=1.0\linewidth]{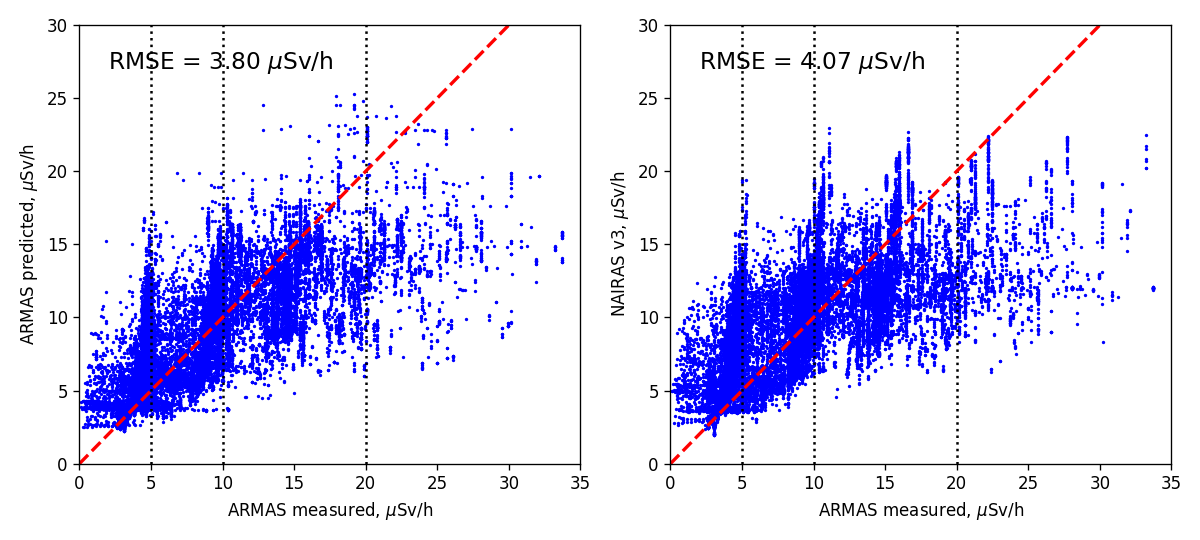}
    \caption{Left: Measured radiation dose rates VS predicted using an ML model (Random Forest Regressor). The ML model was trained on partition 1 of the static dataset and evaluated on partition 2. Right: Measured radiation dose rates VS nowcast of a physics-based NAIRAS~V3 model for partition 2.}
    \label{fig:prediction}
\end{figure}

The result for the case when partition 1 is used for training the ML model, and partition 2 is used for evaluation of its performance and comparison with the predictions of the NAIRAS-v3 physics-based model, is presented in Figure~\ref{fig:prediction}. The left panel illustrates the scatterplot of the radiation dose rates predicted by the ML-driven model against the ARMAS measurements used as a ground truth. One can see that points tend to be mostly organized along the red line, which represents the ideal one-to-one prediction. It is also visible that the predictions demonstrate a stronger spread and systematic deviation from the ideal prediction line for the larger values of dose rates. In fact, the part of the distribution covering the dose rates of $>$15\,$\mu{}Sv/h$ is mostly situated below the perfect prediction line, indicating that the ML-driven predictions typically underestimate the actual dose rates in the cases where the measured radiation doses are high. Interestingly, the comparison of ARMAS measurements with the nowcast of the physics-based NAIRAS-v3 model (presented in the right panel of Figure~\ref{fig:prediction}) demonstrates the same pattern. The root mean squared error (RMSE) computed for the ML-driven prediction is 3.80\,$\mu{}Sv/h$, whereas it is 4.07\,$\mu{}Sv/h$ for the NAIRAS-v3 model. One can see that the ML-driven model seems to deliver slightly more accurate predictions both in terms of the RMSE measure across the test partition and based on the qualitative appearance of the scatterplots in Figure~\ref{fig:prediction} for this partition pair. The situation remains approximately the same for all other partition pairs, with the ratio of RMSEs of the ML-driven radiation nowcasting to NAIRAS-v3-based being equal to $RMSE=0.931\pm{}0.012$ across all six ordered partition pairs.

\begin{figure}
    \centering
    \includegraphics[width=1.0\linewidth]{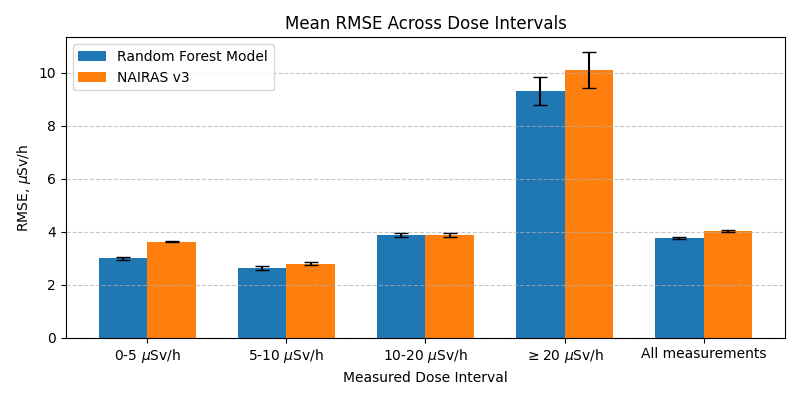} \\
    \includegraphics[width=1.0\linewidth]{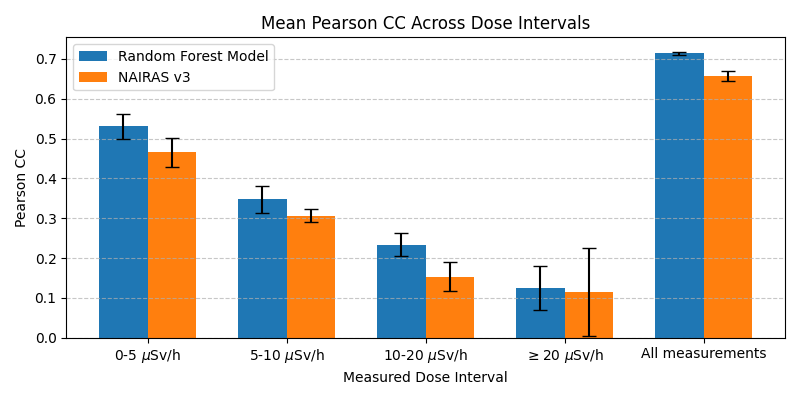}
    \caption{Top: Root mean squared error (RMSE) for the radiation dose rate nowcast obtained with the ML model (Random Forest Regressor, blue) averaged over six possible train-test partition pairs, and the physics-based NAIRAS~V3 model (orange) averaged over three partitions. Bottom: same for the Pearson correlation coefficient.}
    \label{fig:prediction_bins}
\end{figure}

To understand better how the prediction performance varies across the measurements with different radiation intensities, we evaluate the ML-driven and NAIRAS-v3-based nowcasting across all partitions within four ARMAS effective dose rate bins: $\{0-5 \mu{}Sv/h, 5-10 \mu{}Sv/h, 10-20 \mu{}Sv/h, >20 \mu{}Sv/h \}$. Performance is assessed using the RMSE and the linear Pearson correlation coefficient between the predicted and measured radiation dose rates. The results are presented in Figure~\ref{fig:prediction_bins}. It is evident that ML-driven radiation nowcasting model statistically performs better than NAIRAS-v3 model for the effective dose rates of 0-5\,$\mu$Sv/h and $>$20\,$\mu{}Sv/h$. While the improvements are less pronounced for the 5-10\,$\mu$Sv/h and 10-20\,$\mu$Sv/h dose rate bins in terms of RMSE, the related Pearson correlation coefficient is generally higher for the ML-driven models in this range of effective dose rates.

It is also evident that both NAIRAS-v3 and ML models perform very poorly for nowcasting $>20 \mu{}Sv/h$ dose rates. Figure~\ref{fig:prediction_bins} illustrates that the RMSEs for this interval are more than 2 times higher than for the $10-20 \mu{}Sv/h$ dose rate interval, and the corresponding Pearson correlation coefficients are notably small. Accurate predictions for high radiation dose rates are extremely important since these cases can potentially result in significant operational risks and potential radiation impacts. As we can see, both NAIRAS-v3 and ML approaches disagree with ARMAS measurements. However, one can still note a slightly better performance of the ML model, which provides lower RMSEs with respect to NAIRAS-v3, and a stronger correlation with ARMAS measurements. We also would like to note that nowcasting of radiation for high dose rate cases represents a classic imbalance regression challenge: only 4,439 points across 92,476 points in the dataset (about 4.8\%) have the measured ARMAS dose rates of $>$20\,$\mu{}Sv/h$. The corresponding remediation approaches \cite<like those introduced in, e.g., >[]{Chu2025SpWea} might be considered in future efforts.

Overall, this emphasizes the potential for ML approaches in nowcasting the radiation at aviation altitudes. The possible next steps may include generalizing the results over various train-validation-test partition combinations, considering other ML algorithms, and involving the time series data of Geospace parameters for forecast improvement. Our efforts in utilizing other ML algorithms for the presented dataset, along with the initial results of the feature importance analysis, can be found in \citeA{Sanjib2025}. Constructed ML-ready datasets enable such investigations, opening the perspective of the new studies of the research community.

\section{Summary and Discussion}\label{sec:summary}

This paper presents the construction of the ML-ready dataset for nowcasting atmospheric radiation at aviation altitudes. We have leveraged the publicly available effective dose rate measurements by the ARMAS device acquired over 589 flights, and created a pre-partitioned ML-ready dataset, which requires minimal preprocessing to be used for ML purposes. Some of the dataset features are summarized below:
\begin{itemize}
    \item The resulting dataset comprises 589 ARMAS flights containing 92,476 effective dose rate measurements. While the flights are mostly accomplished on top of the continental US and territories (Figure~\ref{fig:armascoverage}), the dataset also samples the regions over the Pacific Ocean, the North Atlantic region, and Antarctica.
    \item The radiation measurements are supported by the measurements from four neutron monitor stations (\texttt{OULU}, \texttt{DOMC}, \texttt{NEWK}, and \texttt{THUL}), solar wind properties at L1, measurements of proton and soft X-ray fluxes by GOES spacecraft series, geomagnetic activity indexes, and global solar activity characteristics (such as sunspot numbers, F10.7 flux, and solar polar fields). This provides an opportunity for comprehensive studies of the dependencies of the radiation environment on the Geospace drivers.
    \item The dataset has been pre-partitioned into three subsets, which can be directly used for training, validation, and testing purposes. During pre-partitioning, it was ensured that the data points from the same flight are within the same partition, and that the sampling of the parameter space is more or less the same within any partition (see Figures~\ref{fig:partitions_primary}~and~\ref{fig:partitions_secondary} for parameter distributions).
    \item Three versions of the dataset are constructed. The `static' version includes the most recent properties of the environment only. The `dynamic' versions include the pre-history of Geospace parameters as time series. The 1-hour and 24-hour long prehistory is considered.
    \item The use case example demonstrates on the selected test subset the possibility of constructing an ML model that predicts the effective dose rates with the average root mean squared error (RMSE) of 3.80$\mu{}$Sv/h, which is slightly better than the NAIRAS v3 physics-based model nowcast (4.07$\mu{}$Sv/h). The results averaged across six possible train-test partition pairs (see Figure~\ref{fig:prediction_bins}) also demonstrate that the ML model typically has a lower RMSE and predicts ARMAS dose rates with a higher Pearson correlation coefficient than the NAIRAS-v3 model. This holds promise for the development of ML-driven models of radiation forecasting in the future.
\end{itemize}

We envision that the constructed datasets could be used for various scientific applications. First, the problem of the atmospheric radiation nowcast given the state of the environment requires a detailed evaluation with respect to the ML algorithms and involvement of the time series. As highlighted in Section~\ref{sec:example}, even the static version of the dataset with the Random Forest regressor demonstrates the promising results. The nowcasting and forecasting results can vary with the algorithm \cite{Ali2024ApJS..270...15A,Goodwin2024ApJ...964..163G,OKeefe2024AdSpR..74.6252O}; therefore, the consideration of other ML methods is necessary. The benefits of including time series properties need to be assessed as well. Second, the slight manipulation of the dataset allows us to consider a forecasting problem instead of nowcasting. All one has to do is consider the `dynamic' dataset versions and avoid considering the data within a certain latency window before the ARMAS measurement. Obviously, the length of the time series (1\,h and 24\,h) would limit the considered latency windows. Third, the dataset could enhance the understanding of radiation environment physics. Besides the standard correlation analyses possible, one could investigate feature importance to understand the influence of some Geospace parameters on the radiation environment \cite<e.g.,>[]{Yeolekar9679962,Sadykov2017ApJ...849..148S}. The developed dataset opens a promising number of prospective studies and facilitates the development of models for the aviation radiation domain.

In addition to subject-related applications, we anticipate that the developed datasets can have broader value with respect to the ML research. As indicated in \citeA{Nita2022,Masson2024}, the ML-ready datasets allow ML practitioners to reduce the time needed for the data preparation phase, and therefore allow them to focus on the development and testing of innovative ML approaches. The developed datasets can serve as a benchmark dataset for both the feature-based and time-series-based forecasting models, providing a unified approach to quantify and compare their performance. Finally, the pronounced imbalanced nature of the radiation data with the limited number of high radiation dose rate measurements, and the challenges the considered models face in predicting these measurements, provide an ideal testbed for ML approaches targeting the imbalanced regression problems.

\section*{Conflict-of-Interest Statement}
The authors have no conflicts of interest to disclose.

\section*{Open Research}
The developed machine learning-ready dataset to nowcast the effective dose rates at aviation altitudes is currently publicly available via the Radiation Data Portal (\url{https://dmlab.cs.gsu.edu/rdp/ml-dataset.html}). The original ARMAS data files are publicly available from the ARMAS Data Archive at Space Environment Technologies (\url{https://sol.spacenvironment.net/ARMAS/Archive/}). The Neutron Monitor data can be accessed via the NMDB database (\url{https://www.nmdb.eu/data/}). The solar wind measurements, geomagnetic activity indexes, and F10.7 index data are publicly accessible via the OMNIWeb service (\url{https://omniweb.gsfc.nasa.gov/ow.html} and \url{https://omniweb.gsfc.nasa.gov/ow_min.html}). The GOES soft X-ray data can be accessed via the National Oceanic and Atmospheric Administration National Centers for Environmental Information Archive (NOAA NCEI, \url{https://www.ncei.noaa.gov/data/goes-space-environment-monitor/access/science/xrs/} and \url{https://data.ngdc.noaa.gov/platforms/solar-space-observing-satellites/goes/}). We thank the developers of the Integrated Space Weather Analysis (ISWA, \url{https://ccmc.gsfc.nasa.gov/tools/ISWA/}) systems API for the possibility of retrieving the operational GOES integrated proton fluxes. The daily sunspot number is obtained via the publicly accessible World Data Center SILSO (\url{https://www.sidc.be/SILSO/datafiles}). The solar polar field measurements are publicly available via the Wilcox Solar Observatory website (\url{http://wso.stanford.edu/Polar.html}).

\appendix

\section{Description of ML-Ready Data Set Features}

The features that are presented in the ML-ready dataset are summarized in Table~\ref{tab:dataset_properties}. Please note that the ``Cadence used'' column indicates the cadence of the data that has been used for the dataset construction, and that the actual cadence of the corresponding instrument measurements can be higher. The abbreviation GSM refers to Geocentric Solar Magnetospheric coordinate system. The units of $pfu$ for energetic particle measurements are 1 pfu $=$ 1 particle$\cdot{}cm^{-2}\cdot{}s^{-1}$, the units of sfu used for f10.7 flux measurement are 1 sfu $=$ 10$^{-22}$W$\cdot$m$^{-2}\cdot$Hz$^{-1}$. We also provide some additional information regarding the data sources used for the dataset construction below:

The pressure- and efficiency-corrected neutron monitor measurements can be accessed via the NMDB database (\url{https://www.nmdb.eu/}). More information about the data in NMDB can be found here: \url{https://www.nmdb.eu/nest/help.php#helptable/}. The detailed description of the low-resolution (1-hour) OMNIWeb dataset can be found at \url{https://omniweb.gsfc.nasa.gov/ow.html}. For high-resolution (5-minute) OMNIWeb dataset, one can find the details at \url{https://omniweb.gsfc.nasa.gov/ow_min.html}. The details on the SILSO sunspot number series can be found in \url{https://doi.org/10.24414/qnza-ac80}. The soft X-ray measurements of the science-quality 1-minute cadence data have been downloaded from the National Centers for Environmental Information archive. For the data prior to 2017-02-07, the data has been obtained from \url{https://www.ncei.noaa.gov/data/goes-space-environment-monitor/access/science/xrs/}. For further dates, the level-2 data from \url{https://data.ngdc.noaa.gov/platforms/solar-space-observing-satellites/goes/} is used.

The near-real-time energetic particle data (for energetic electrons and protons) has been obtained using the Integrated Space Weather Analysis (ISWA, \url{https://ccmc.gsfc.nasa.gov/tools/ISWA/}) systems API that collects the data directly from Space Weather Prediction Center's data service (\url{https://services.swpc.noaa.gov/json/goes/}). The $E\geq{}2$\,MeV 5-minute energetic electron flux has been corrected to some extent with respect to the galactic cosmic ray background and solar proton contamination. The description of the correction algorithm for the Energetic Proton, Electron and Alpha Detectors (EPEAD) instrument onboard GOES 13-15 can be found in \citeA{Rodriguez2014_GOES-EPEAD}. For the details of the cross-calibration between the EPEAD and Magnetospheric Particle Sensor - High Energy (MPI-HI) onboard GOES 16 and following can be found in \citeA{Boudouridis2020SpWea..1802403B}. The integral fluxes of the energetic proton data represent the derived quantities. For the integral flux derivation procedures, please see the Appendix (with respect to EPEAD onboard GOES 13-15) and Section 3 (with respect to Solar and Galactic Proton Sensor, SGPS, onboard GOES 16-19) of \citeA{Rodriguez2017SpWea..15..290R}. For the cross-calibration of SGPS onboard GOES-16 and EPEAD onboard GOES-13 and GOES-15 during September 2017 events, please see \citeA{Kress2021SpWea..1902750K}.

\begin{center}

\begin{longtable}{|p{0.17in}|p{1.4in}|p{1.9in}|p{0.8in}|p{0.6in}|}
\caption{Summary of features and measurement properties used for the construction of the machine learning ready datasets for aviation radiation nowcasting.}
\label{tab:features} \\
\hline
\rowcolor[gray]{0.9}
\textbf{No.} & \textbf{Feature ID} & \textbf{Feature meaning} & \textbf{Units /  Format} & \textbf{Cadence Used} \\
\hline
\endfirsthead

\hline
\rowcolor[gray]{0.9}
\textbf{No.} & \textbf{Feature ID} & \textbf{Feature meaning} & \textbf{Units / Format} & \textbf{Cadence Used} \\
\hline
\endhead

\endfoot
\hline
\endlastfoot

\rowcolor[gray]{0.85}
\multicolumn{5}{|c|}{\textbf{Flight and Dosimetry Properties}} \\ \hline

1  & Datetime               & Time of ARMAS measurement             & yyyy-mm-dd HH:MM:SS & 1 min \\ \hline
2  & ARMAS                  & ARMAS derived effective dose rate     & $\mu$Sv/h           & 1 min \\ \hline
3  & NAIRASV3               & NAIRAS v3 modeled effective dose rate & $\mu$Sv/h           & 1 min \\ \hline
4  & NAIRASV2               & NAIRAS v2 modeled effective dose rate & $\mu$Sv/h           & 1 min \\ \hline
5  & Latitude               & Geographic latitude                   & deg                 & 1 min \\ \hline
6  & Longitude              & Geographic longitude                  & deg                 & 1 min \\ \hline
7  & Altitude (Bar)         & Barometric altitude                   & km                  & 1 min \\ \hline
8  & Altitude (GPS)         & GPS altitude                          & km                  & 1 min \\ \hline
9  & Geomagnetic latitude   & Geomagnetic latitude                  & deg                 & 1 min \\ \hline
10 & Geomagnetic longitude  & Geomagnetic longitude                 & deg                 & 1 min \\ \hline
11 & Geomagnetic Rc         & Geomagnetic cutoff rigidity           & GV                  & 1 min \\ \hline
12 & Geomagnetic L-shell    & L-shell                               & --                  & 1 min \\ \hline
13 & Vehicle ID             & Aircraft identifier                   & --                  & Per flight \\ \hline

\rowcolor[gray]{0.85}
\multicolumn{5}{|c|}{\textbf{Neutron Monitor Measurements (pressure- and efficiency-corrected)}} \\ \hline

14 & NM\_NEWK & NEWK station count rate & counts/s & 5 min \\ \hline
15 & NM\_OULU & OULU station count rate & counts/s & 5 min \\ \hline
16 & NM\_THUL & THUL station count rate & counts/s & 5 min \\ \hline
17 & NM\_SOPO & SOPO station count rate & counts/s & 5 min \\ \hline

\rowcolor[gray]{0.85}
\multicolumn{5}{|c|}{\textbf{Soft X-ray Flux Measurements}} \\ \hline

18 & SXR\_short & 0.5--4 \AA{} flux & W·m$^{-2}$ & 1 min \\ \hline
19 & SXR\_long  & 1--8 \AA{} flux   & W·m$^{-2}$ & 1 min \\ \hline

\rowcolor[gray]{0.85}
\multicolumn{5}{|c|}{\textbf{Energetic Particle Measurements}} \\ \hline

20 & Particles\_P1   & Proton flux, $E \geq 1$ MeV     & pfu & 5 min \\ \hline
21 & Particles\_P5   & Proton flux, $E \geq 5$ MeV     & pfu & 5 min \\ \hline
22 & Particles\_P10  & Proton flux, $E \geq 10$ MeV    & pfu & 5 min \\ \hline
23 & Particles\_P30  & Proton flux, $E \geq 30$ MeV    & pfu & 5 min \\ \hline
24 & Particles\_P50  & Proton flux, $E \geq 50$ MeV    & pfu & 5 min \\ \hline
25 & Particles\_P100 & Proton flux, $E \geq 100$ MeV   & pfu & 5 min \\ \hline
26 & Particles\_E20  & Electron flux, $E \geq 2$ MeV   & pfu & 5 min \\ \hline

\rowcolor[gray]{0.85}
\multicolumn{5}{|c|}{\textbf{Solar Wind Measurements (GSM coordinates where applicable)}} \\ \hline

27 & SW\_B           & Magnetic field magnitude         & nT & 5 min \\ \hline
28 & SW\_Bx          & $B_x$ component            & nT & 5 min \\ \hline
29 & SW\_By          & $B_y$ component            & nT & 5 min \\ \hline
30 & SW\_Bz          & $B_z$ component            & nT & 5 min \\ \hline
31 & SW\_V           & Velocity magnitude               & km/s & 5 min \\ \hline
32 & SW\_Vx          & $V_x$ component            & km/s & 5 min \\ \hline
33 & SW\_Vy          & $V_y$ component            & km/s & 5 min \\ \hline
34 & SW\_Vz          & $V_z$ component            & km/s & 5 min \\ \hline
35 & SW\_density     & Proton number density                   & cm$^{-3}$ & 5 min \\ \hline
36 & SW\_temperature & Proton temperature                      & K & 5 min \\ \hline
37 & SW\_pressure    & Flow pressure                 & nPa & 5 min \\ \hline

\rowcolor[gray]{0.85}
\multicolumn{5}{|c|}{\textbf{Geomagnetic Activity Indexes}} \\ \hline

38 & Index\_Kp  & Planetary Kp index            & -- & 1 h \\ \hline
39 & Index\_Dst & Disturbance storm-time index  & nT & 1 h \\ \hline
40 & Index\_Ap  & Planetary Ap index            & nT & 1 h \\ \hline

\rowcolor[gray]{0.85}
\multicolumn{5}{|c|}{\textbf{Global Solar Activity Properties}} \\ \hline

41 & Solar\_sunspots & Daily sunspot number                  & --     & 1 day      \\ \hline
42 & Solar\_f107     & Solar f10.7 index                     & sfu     & 1 h      \\ \hline
43 & Solar\_NPF      & North polar magnetic field            & $\mu$T & 10 days \\ \hline
44 & Solar\_SPF      & South polar magnetic field            & $\mu$T & 10 days \\ \hline
45 & Solar\_APF      & Average polar magnetic field          & $\mu$T & 10 days \\ \hline
46 & Solar\_NPF20    & 20\,nHz-filtered north polar magnetic field    & $\mu$T & 10 days \\ \hline
47 & Solar\_SPF20    & 20\,nHz-filtered south polar magnetic field   & $\mu$T & 10 days \\ \hline
48 & Solar\_APF20    & 20\,nHz-filtered average polar magnetic field & $\mu$T & 10 days
\label{tab:dataset_properties}
\end{longtable}

\end{center}

\acknowledgments
This work has been supported by the NASA HITS grant 80NSSC22K1561. VMS also thanks the NSF FDSS grant 1936361 and NASA LWS grant 80NSSC24K1111. We acknowledge the NMDB database (\url{www.nmdb.eu}), founded under the European Union's FP7 programme (contract no. 213007) for providing data, and the PIs of individual neutron monitors at: Newark and Thule (University of Delaware Department of Physics and Astronomy and the Bartol Research Institute, USA), Kerguelen (Observatoire de Paris and the French Polar Institute IPEV, France), Oulu (Sodankyla Geophysical Observatory of the University of Oulu, Finland), South Pole (University of Wisconsin, River Falls, USA). We acknowledge SILSO, OMNIWeb, Wilcox Observatory, ISWA services, and SWPC/NOAA for the availability of the high-quality data used in the dataset construction and for their clarifications regarding observational data details. We would also like to thank Juan Rodriguez (CU Boulder, CIRES) for invaluable insights with respect to the GOES energetic particle data used in this work.

\bibliography{ARMAS_ML}

\end{document}